\RequirePackage[2020-02-02]{latexrelease}
\documentclass[twocolumn,amsmath,amssymb]{snp}
\usepackage[switch]{lineno} 
\usepackage{lipsum} 
\usepackage{graphicx}
\usepackage{dcolumn}
\usepackage{bm}
\usepackage{subfigure}
\begin{document}

\title{ \Large {Study of $\pi$, K, and p production in high multiplicity pp collisions at $\sqrt{s}$ = 13 TeV with ALICE at the LHC}}
\author{\large Navneet Kumar ``for the ALICE Collaboration''}
\email{navneet.kumar@cern.ch}
\affiliation{ Department of Physics, Panjab University, Chandigarh, 160014,  INDIA}


\maketitle
\section*{Introduction}
The primary objective of the heavy-ion physics is to investigate the characteristics of a state of matter known as the quark--gluon plasma (QGP), where the quarks and gluons are no longer confined into hadrons and chiral symmetry is restored. The CERN Large Hadron collider (LHC) demonstrated that in high-energy heavy-ion (A--A) collisions, a strongly interacting QGP is formed~\cite{dae1}. Recent reports on the enhancement of (multi-)strange hadrons, double-ridge structure, non-zero $v_{2}$ coefficient, mass ordering in hadron $p_{\rm T}$ spectra,  and modifications of baryon-to-meson ratios suggested that the collective phenomena (signatures of QGP formation) is also present in the p--Pb collisions~\cite{dae2}. These studies are further extended to smaller systems, such as pp collisions where similar observations have been reported in the high-multiplicity events. These high multiplicity events are categorized by the charged-particle multiplicity at midrapidity region ${\langle {\rm d}N_{\rm ch}/{\rm d}\eta \rangle}_{|\eta|<0.5}$ across different systems and collision energies. The measurement of transverse momentum ($p_{\rm T}$) spectra and yield of identified particles unveil the properties of matter created in the collisions e.g. strangeness enhancement and radial flow~\cite{dae3,dae4,dae5}. The $p_{\rm T}$-integrated particle yield is sensitive to the production mechanism~\cite{dae6} and K/$\pi$ ratio as a function of multiplicity indicates enhanced strangeness production~\cite{dae2}. Thus the systematic investigation of identified particle production might lead to a better understanding of the collision dynamics in small and large systems. In this contribution, we will present the recent results on $\pi$, K, and p in the highest multiplicity classes: 0.00$\%$-0.01$\%$, 0.01$\%$-0.05$\%$, 0.00$\%$-0.10$\%$ for pp collisions at $\linebreak$$\sqrt{s}$ = 13 TeV. Furthermore, the comparison of these results with published results will be presented.

\section*{Analysis details}
This analysis is carried out on the data set of pp collisions with an energy of $\linebreak$$\sqrt{s}$ = 13 TeV collected in 2016 and 2018. The V0 detector covering the pseudo-rapidity range  $\linebreak$-3.7 $< \eta <$  -1.7 and 2.8 $< \eta <$ 5.1 has been used for classifying events by measuring the charged-particle multiplicity (${\rm d}N_{ch}/{\rm d}\eta $). All the events considered in this analysis are recorded using the high-multiplicity trigger and the reconstructed primary vertex is required to be located within 10 cm from the nominal interaction point of the ALICE apparatus. 
\par The particle identification is performed with the help of several ALICE sub-detectors, including the Inner Tracking System (ITS), Time Projection Chamber (TPC), Time-of-Flight detector (TOF). Furthermore, the charged pions and kaons have been identified using a topological decay (kinks) approach within the TPC volume. 
\section{Results and discussion}
\begin{figure}[h]

  \includegraphics[width=60mm]{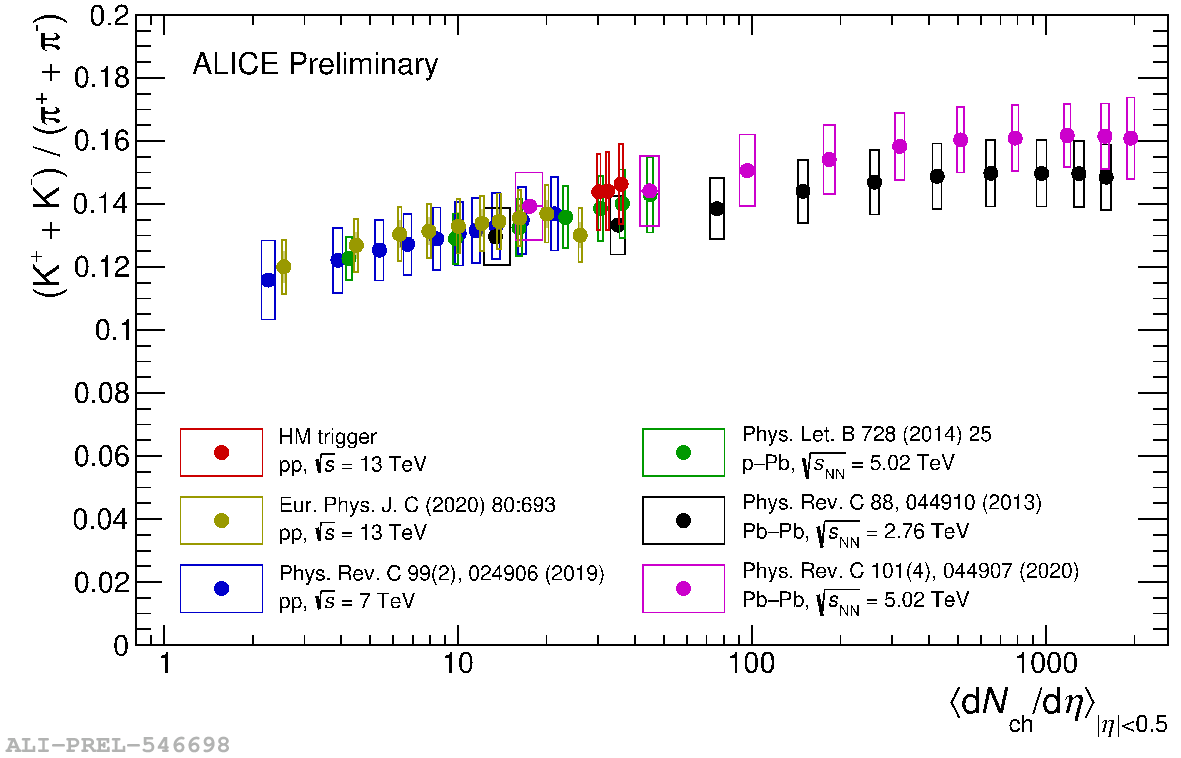}
  \caption{\label{fig1} K/$\pi$ ratios as a function of $\langle {\rm d}N_{\rm ch}/{\rm d}\eta \rangle$ for different collision systems and energies.}
\end{figure}
\begin{figure}[h]
  
  \includegraphics[width=60mm]{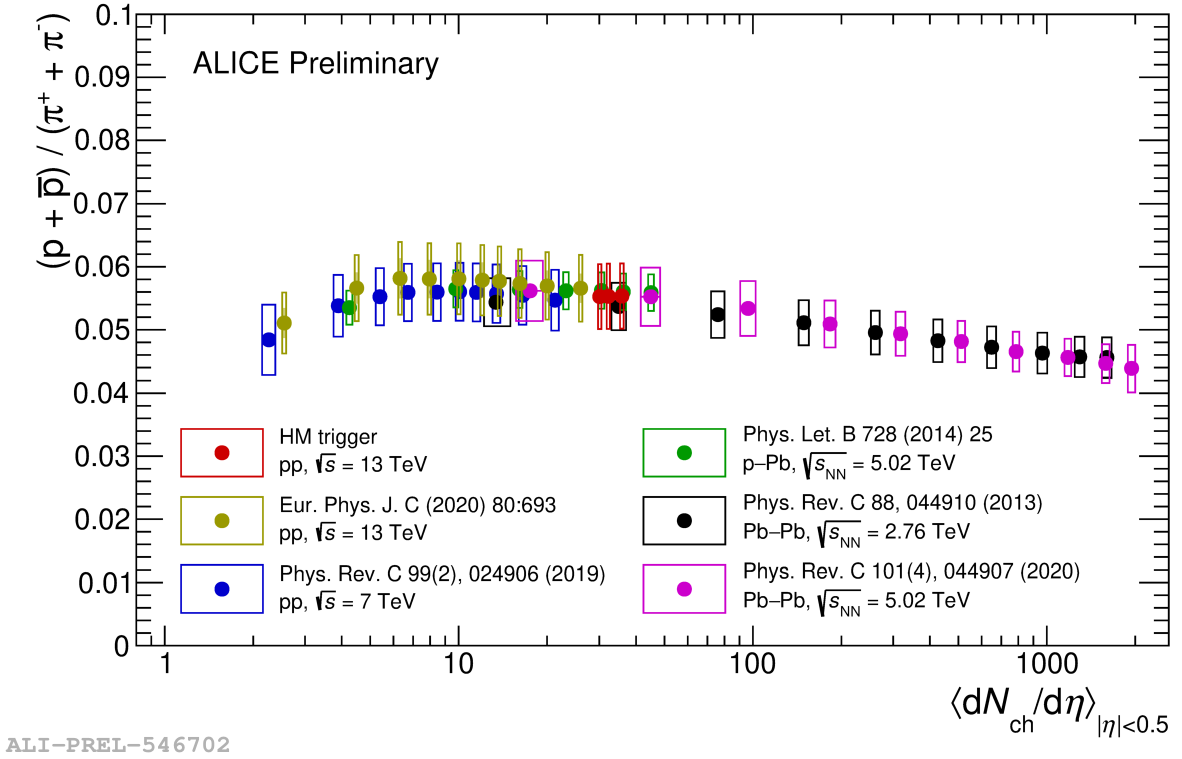}
  \caption{\label{fig2} p/$\pi$ ratios as a function of $\langle {\rm d}N_{\rm ch}/{\rm d}\eta \rangle$ for different collision systems and energies.}
\end{figure}
\begin{figure}[h! ]
  \includegraphics[width=70mm, height =35mm]{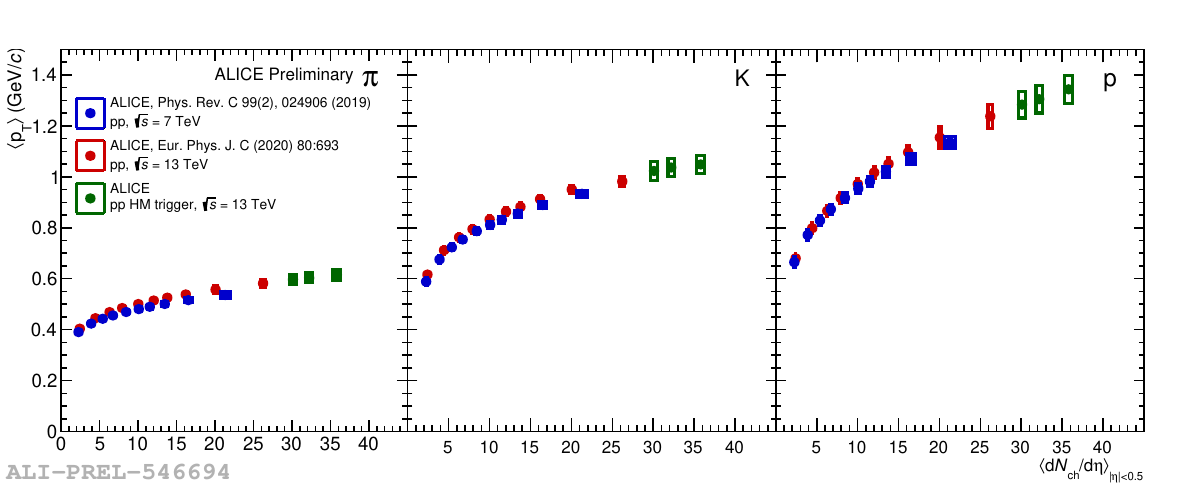}
  \caption{\label{fig3}${\langle p_{\rm T} \rangle}$ of $\pi$ (left), K (mid), and p (right) as a function of $\langle {\rm d}N_{\rm ch}/{\rm d}\eta \rangle$ for pp collisions at $\sqrt{s}$ = 7 and 13 TeV.}
\end{figure}

In Fig.~\ref{fig1} and~\ref{fig2}, the $p_{\rm T}$-integrated K/$\pi$ and p/$\pi$ ratios are plotted as a function of multiplicity for different energies and colliding systems. The K/$\pi$ yield ratio increases from low multiplicity pp events to central Pb--Pb events, this increase is attributed to the enhanced production of strangeness at larger freeze-out volumes while the p/$\pi$ ratio decreases continuously after $\langle {\rm d}N_{\rm ch}/{\rm d}\eta \rangle >100$. The center of mass energy and colliding system have no influence on the the  p/$\pi$ and K/$\pi$ ratios. 
\par Fig.~\ref{fig3} shows the ${\langle p_{\rm T} \rangle}$ for charged pions, kaons and (anti-)protons as a function of the charged particle multiplicity density  $\langle {\rm d}N_{\rm ch}/{\rm d}\eta \rangle$ at midrapidity in pp collisions at $\sqrt{s}$ = 7 and 13 TeV. The average transverse momentum of identified particles is found to increase with multiplicity in pp collisions as found in Pb--Pb collisions and this rise of average $p_{\rm T}$ gets steeper with increasing hadron mass, this effect is consistent with the presence of radial flow~\cite{dae3,dae4,dae5}.

\section*{Acknowledgments}
The author acknowledge financial support from Council of Scientific $\&$ Industrial Research (CSIR), and DST-project grant No. SR/MF/PS-02/2021-PU India.



\end{document}